\documentclass[10pt,conference]{IEEEtran}
\IEEEoverridecommandlockouts
\usepackage{amsmath,amssymb,amsfonts}
\usepackage{algorithmic}
\usepackage{graphicx}
\usepackage{textcomp}
\usepackage{xcolor}
\usepackage{tabu}
\usepackage[backend=biber,uniquelist=false,maxbibnames=9,maxcitenames=2,mincitenames=1,style=ieee, 
url=false, isbn=false, eprint=false,
doi=false,
natbib=true]{biblatex}
\usepackage{multirow}
\usepackage[inline]{enumitem}
\usepackage{graphicx}
\usepackage{gensymb}
\usepackage{url}
\usepackage{subcaption}
\usepackage{hyphenat}
\usepackage[font=footnotesize]{caption}
\def\*#1{\mathbf{#1}}
\def\!#1{\mathbb{#1}}
\def\.#1{\mathcal{#1}}

\definecolor{red}{RGB}{0,0,0}

\begin{document}

\title{Speaker Embeddings to Improve Tracking of Intermittent and Moving Speakers}
% SPEAKER EMBEDDINGS TO IMPROVE TRACKING OF INTERMITTENT AND MOVING SPEAKERS

\author{
\IEEEauthorblockN{Taous Iatariene$^{12}$, Can Cui$^2$, Alexandre Guérin$^1$, Romain Serizel$^2$}

\IEEEauthorblockA{
\textit{$^1$
 Orange Innovation}, Rennes, France \\
\textit{$^2$
 University de Lorraine, CNRS, Inria, Loria}, Nancy, France \\
}

taous.iatariene@orange.com, can.cui@inria.fr, alexandre.guerin@orange.com, romain.serizel@loria.fr
}

\maketitle

\begin{abstract}

Speaker tracking methods often rely on spatial observations to assign coherent track identities over time.
This raises limits in scenarios with intermittent and moving speakers, i.e., speakers that may change position when they are inactive, thus leading to discontinuous spatial trajectories.
This paper \textcolor{red}{investigates} the use of speaker embeddings, {
% as a first step towards a simple yet effective solution.
in a simple solution 
to this issue}.
We propose to perform identity reassignment {post-tracking}, using speaker embeddings.
We leverage trajectory-related information provided by an initial tracking step and \textcolor{red}{the} multichannel audio signal. Beamforming is used to enhance the signal towards the speakers' positions in order to compute speaker embeddings. These are then used to assign new track identities based on {an enrollment pool}.
We evaluate the performance of the proposed speaker embedding-based identity reassignment method on a dataset where speakers change position
% of speakers changing position 
during inactivity periods.
Results show that it consistently improves the identity assignment performance 
% an exhaustive panel of 
of neural and standard tracking systems.
In particular, we study the impact of beamforming and input duration for embedding extraction.

\end{abstract}


\section{Introduction}

Knowing the position of the speakers present in an acoustic scene is useful in many applications, such as 
speech enhancement~\cite{xenakiSoundSourceLocalization2018}
%, speech separation~\cite{lluisDirectionSpecificAmbisonics2023}, 
and automatic speech recognition~\cite{subramanianDeepLearningBased2022}.
For static speakers, this involves localization, which provides estimations of the speakers' Directions of Arrival (DoAs). 
In the presence of multiple and moving speakers, an additional step of tracking is necessary to link the DoAs through time and assign them coherent track identities. 

Many studies employ a Bayesian filtering framework for tracking~\cite{quinlanTrackingIntermittentlySpeaking2009,fallonAcousticSourceLocalization2012,
liOnlineLocalizationTracking2019, croccoAudioTrackingNoisy2018a,hoggMultichannelOverlappingSpeaker2021} where data association techniques are used to deal with multiple sources and track identity management~\cite{voMultitargetTracking2015}. 
Recently, deep learning methods were proposed, in which neural networks are trained to perform ordered localization, using permutation invariant loss functions~\cite{adavanneDifferentiableTrackingBasedTraining2021, diaz-guerraPositionTrackingVarying2023a, grinsteinNeuralSRPMethodUniversal2024a}.

% Significant challenges can be encountered when facing 
% dynamic scenarios, such as the presence of 
Tracking intermittent and moving speakers, who may change position when silent, is a seldom explored problem~\cite{quinlanTrackingIntermittentlySpeaking2009,fallonAcousticSourceLocalization2012,liOnlineLocalizationTracking2019}.
~\citet{liOnlineLocalizationTracking2019}
proposed to track both active and silent speakers and 
assumed spatial continuity to update inactive speaker's trajectories. 
\textcolor{red}{This raised limits when spatial continuity was not ensured, i.e, when speakers moved unpredictably while silent, which lead to discontinuous spatial trajectories~\cite{iatarieneTrackingIntermittentMoving2025}.}
% occurring into speakers changing directions during silences, thus leading to discontinuous spatial trajectories.
% \textcolor{red}{While this proved to be efficient under the assumption of }

% {It raises challenges due to the presence of discontinuous spatial trajectories
% ~\cite{liOnlineLocalizationTracking2019}} associates a weight (prob that a spk is active) for each spatial observation / assotation-to-track becomes a latent variable / track a max number of speaker N simultaneously whether active or not / speaker birth process + speech activity detection.
% \cite{quinlanTrackingIntermittentlySpeaking2009} : track max number of speakers N simultaneously whether active or not / estimate nos Na and keep the Na most active tracks.
% \cite{fallonAcousticSourceLocalization2012}
% To maintain coherent identity assignment over time in such cases,
% This led~\citet{quinlanTrackingIntermittentlySpeaking2009}~to propose to explicitly identify the active speakers while performing bayesian filtering-based tracking.}

Another solution for tracking intermittent and moving speakers would be to combine spatial and identity-related observations. 
% Most speaker tracking systems rely on DoAs (i.e., spatial observations), to reconstruct speaker trajectories. 
% Still, some studies investigated the combination of DoAs and identity-related observations for tracking such as 
Spectral signatures~\cite{croccoAudioTrackingNoisy2018a} and speaker's fundamental frequencies~\cite{hoggMultichannelOverlappingSpeaker2021} have been investigated for speaker tracking, but never for tracking intermittent and moving speakers.

% Speaker-specific information can be extracted using neural networks trained 
% ~\cite{snyderXVectorsRobustDNN2018, desplanquesECAPATDNNEmphasizedChannel2020}.
Speaker embeddings are identity-related observations obtained through recent advances in deep learning~\cite{snyderXVectorsRobustDNN2018}, which have gained popularity due to their superior ability to distinguish between speakers, compared to other speaker-related features~\cite{croccoAudioTrackingNoisy2018a,hoggMultichannelOverlappingSpeaker2021}.
They are commonly used in speaker
% identification~\cite{hongCombiningDeepEmbeddings2020}, 
verification~\cite{desplanquesECAPATDNNEmphasizedChannel2020,jakubecDeepSpeakerEmbeddings2024} and diarization~\cite{parkReviewSpeakerDiarization2022}.
To the best of our knowledge, they have never been used to complement DoAs for speaker tracking.

% The proposed solution to this rarely addressed issue is novel in the following ways :
% \begin{enumerate*}
%     \item It uses speaker embeddings as identity-related observations for tracking.
%     \item It combines spatial and identity-related observation for tracking intermittent and moving speakers.
% \end{enumerate*}

This paper proposes to combine DoAs and 
speaker embeddings as 
identity-related observations, to correct the assignment errors caused by the discontinuous spatial trajectories of speakers changing position during inactivity periods.
% , thus improving tracking of intermittent and moving speakers.
The provided solution performs identity reassignment over the output of a tracking system, given an enrollment pool of embeddings.
To reduce the impact of noise and overlapping speakers for embedding extraction, beamforming is applied beforehand, that leverages the DoA information provided by the tracking system's spatial trajectories.

% We evaluate our solution on a dataset containing intermittent and moving speakers. 
We evaluate the global performance of the proposed method for different enrollment pool sizes. We show that it allows for improvements in terms of identity assignment, proving that speakers embeddings can in fact be interesting identity-related observations for tracking intermittent and moving speakers.
We study several factors impacting speaker embedding quality, namely, the impact of input signal length for extraction, as well as the impact of beamforming. 
We finally assess the robustness of \textcolor{red}{the} post-tracking method against several tracking systems, by studying the impact of spatial trajectory quality on the performance. 

% {[Phrase short context / towards low latency + phrase results]}

% {This paper is organized as follows. Section~\ref{sec:problem} presents the problem and introduces our approach. Section~\ref{sec:exp} describes the experimental setup and section~\ref{sec:res} presents the results and their analysis. We conclude in Section~\ref{sec:conclu}.}

\begin{figure*}[!htbp]
    \centering
    \includegraphics[width=0.95\linewidth]{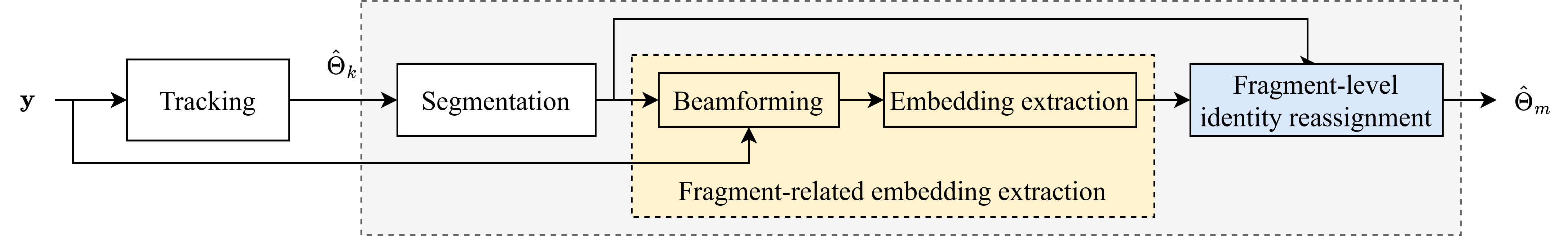}
    \caption{Overview of the proposed speaker embedding-based reassignment method.}
    \label{fig:system}
\end{figure*}

\section{Proposed Method}
\label{sec:problem}

\subsection{Problem formulation}
 % the goal of a speaker tracking system is to generate a unique trajectory per speaker.
We consider a multichannel audio mixture $\*y$ composed of $J$ speakers in a reverberant environment:
\begin{equation}
\label{eq:mixture}
    \*y = \sum_{j=1}^{J} \*{x}_j + \*{n}
\end{equation}
where $\*n$ is a diffuse noise.
{Each speaker, in addition to its clean speech $\*x_j$, 
% with $j \in \{ 1, ..., J\}$, 
can be described by its spatial trajectory $\Theta_j$, which is the set of DoAs associated over the entire duration of the mixture recording.}

A tracking system can predict $K$ trajectories $\hat{\Theta}_k$ with $k \in \{ 1, ..., K\}$. 
\textcolor{red}{We define a fragment as a period of activity within a trajectory. }
{Unlike a trajectory, which may include DoA discontinuities due to track inactivity, a fragment represents a continuous period of activity 
% which is assumed to belong to a single speaker.
assigned by the tracker to a single identity, 
with no DoA interruption.}

% RS : Préciser que le locuteur est actif de manière continue sur un fragment (pas sur une trajectoire) et qu'il n'y a pas de discontinuité de DoA sur les fragments (alors que ça peut être le cas dans les trajectoires.
% AG ; préciser qu'un fragment est associé à une source, donc à une période d'activité continue d'une source en particulier

{We consider scenarios with intermittent and moving speakers.  Movements occurring during speaker activity can be managed with common tracking systems, for example by relying on data association techniques \cite{voMultitargetTracking2015}.
We therefore choose to discard them and focus on speakers whose movements occur only during inactivity periods. 
% may change position during periods of inactivity but remain stationary while speaking.
}
% RS: expliquer que le problème des sources qui bougent pendant les périodes d'activités est étudié par ailleurs et peut être résolu en s'appuyant sur la continuité spatiale à court termes. C'est pourquoi on choisi de se concernter sur les sources qui bougent pendant les périodes d'inactivité.
% The movements occurring during silence 
Such movements disrupt the spatial trajectory continuity, which the tracker relies on for identity assignment. 
This leads to potential errors like assigning multiple identities to the same speaker~\cite{liOnlineLocalizationTracking2019}.

\subsection{\nohyphens{Fragment-level identity reassignment using speaker emebddings}}

To address the aforementioned issue, 
% which cannot be resolved using spatial information only, 
we propose
% a first solution 
% relying on speaker embeddings 
% correcting the identity assignment errors of a speaker tracking system.
a speaker embedding-based identity reassignment method, for which an overview is given in Fig.~\ref{fig:system}.
Given a tracking system, 
% each trajectory $\hat{\Theta}_k$ is segmented into fragments. 
our proposed method predicts $M$ new trajectories $\hat{\Theta}_m$ with $m \in \{ 1, ...,M\}$, by computing fragment-related speaker embeddings, then performing fragment-level identity reassignment, i.e., assigning new identity to each trajectory fragment.
% We rely on the assumption that the tracking system makes little to no errors in the intra-fragment level.

{Each trajectory $\hat{\Theta}_k$ is first segmented into fragments.}
{Robust fragment-related speaker embeddings are extracted with a model trained for speaker identification on a large scale dataset containing a high variety of speakers.}
Due to the presence of noise and overlapping speech induced by the presence of multiple speakers, beamforming is applied to the multichannel mixture $\*y$ before embedding extraction.
This spatial filtering technique can leverage the fragment DoA information and enhance the signal towards the fragment's direction, to extract more robust speaker embeddings. 

% {In multi-speaker scenarios, overlapping speech is inevitable, and encoding speaker information from overlapped speech may degrade speaker representativity. Therefore, we chose to perform beamforming to extract separated speech. Since the speakers' location information is already provided by the DoA-based tracker, we use this information to perform spatial beamforming for each fragment.}

To perform identity reassignment, a pool of enrollment embeddings is computed beforehand. The number of enrollments determines the maximum number of identities our system can predict, i.e. $M$.
We consider two cases: the first case with $M=J$ enrollment embeddings corresponds to an enrollment at the beginning of a session; the second case $M > J$ assumes that the speakers present in the sessions are unknown beforehand but belong to a predefined pool.
% (e.g., employees enrolled once for future enrollment).

{Since one of our main focuses is to assess the interest of speaker embeddings as identity-related observation for tracking}, we choose a naïve first-in-first-out method to perform fragment-level identity reassignment, in which fragments are reassigned by temporal order of appearance.
The new fragment identity is attributed by selecting the identity of the enrollment embedding that has the highest cosine similarity score with the corresponding fragment-related speaker embedding. Since the same identity cannot be assigned to two overlapping fragments, enrollment embeddings of previously assigned and overlapping fragments are discarded.

\section{Experimental Design}
\label{sec:exp}

% Evaluating the improvement in terms of re-identification can only be done in acoustic scenes with dynamic speakers.

% , who are static during speech activity periods but can change position during inactivity periods.

\subsection{Datasets}
\label{sec:datasets}
\textcolor{red}{Experiments are conducted on the LibriJump-2spk dataset~\cite{iatarieneTrackingIntermittentMoving2025}, which is an evaluation dataset of 150 simulated acoustic scenes in the First Order Ambisonics (FOA) format, each containing $J=2$ intermittent and moving speakers from the LibriSpeech~\cite{panayotovLibrispeechASRCorpus2015} test-clean subset. While in LibriJump-2spk the speakers are spaced with angular distances of at least $60\degree$, we further add experiments with 150 other scenes containing closer speakers ($25-60\degree$). The SNR is fixed to 15~dB, and a 2-4~dB level difference is randomly set between the two speakers.}

\textcolor{red}{As detailed by~\citet{iatarieneTrackingIntermittentMoving2025}, movement during silence is simulated by convolving each voice activity period from each speaker, with a unique Spatial Room Impulse Response (SRIR) from the same room but at a different location. 
}

{
For localization and tracking, training datasets are constructed similarly to  LibriJump acoustic scenes generation~\cite{iatarieneTrackingIntermittentMoving2025}, except for the following differences :
\begin{itemize*}
    \item Separate training subsets are used for SRIRs, noise samples, and speech utterances (Librispeech train-clean-100 and train-clean-360).
    \item SNR varies randomly between 5 and 50 dB.
    \item Speakers are static and do not move when silent.
    Movement is indeed unnecessary due to the models' short input sequence lengths~\cite{grumiauxImprovedFeatureExtraction2021}, over which a speaker can be considered as static.
\end{itemize*}
}

\subsection{Tracking systems}

We propose three tracking systems to assess the robustness of our fragment-level identity reassignment system, forming an exhaustive panel of current 
% state-of-the-art 
tracking approaches:
\begin{enumerate*}
    \item \textbf{GT}: Bayesian tracking based on ground truth DoAs.
    \item \textbf{EST}: Bayesian tracking based on estimated DoAs.    
    \item \textbf{NN}: Neural tracking through the prediction of ordered DoAs.
\end{enumerate*}

The Bayesian tracker is a particle filter~\cite{kiticTRAMPTrackingRealtime2018}, in which the maximum number of identities predicted is fixed to the number of enrollment embeddings $M$ used in the experiments. 
{The probability of new source appearance is changed to adapt the nature of the observations (ground truth or estimated).
}

Estimated DoAs are obtained through a Convolutional Recurrent Neural Network (CRNN)~\cite{grumiauxImprovedFeatureExtraction2021}.
The neural tracker is an adaptation of the source splitting approach~\cite{subramanianDeepLearningBased2022} to the FOA multichannel format. It relies on Permutation Invariant Training (PIT) to predict ordered DoAs. The number of identities this neural tracker can predict is fixed by the number of output branches in the architecture (in our case, it is the number of speakers in the scenes, so 2).

\subsection{Speaker Embedding Extraction}

The state-of-the-art ECAPA-TDNN~\cite{desplanquesECAPATDNNEmphasizedChannel2020} model is used for speaker embedding extraction in our experiments. It takes as input MFCCs of a speech signal of variable length, to output a latent representation of the speaker, encoded in a vector of dimension 192.
We use in particular a pre-trained and open-source version developed by speechbrain~\cite{ravanelliSpeechBrainGeneralPurposeSpeech2021} which have been trained on the Voxceleb1 and Voxceleb2 datasets~\cite{nagraniVoxCelebLargeScaleSpeaker2017} for a speaker identification task\footnote{\url{https://huggingface.co/speechbrain/spkrec-ecapa-voxceleb}}.

Our method relies on fragment-related speaker embeddings, as well as enrollment embeddings for fragment-level identity reassignment. 
One \textcolor{red}{robust} enrollment embedding is extracted per speaker present in the considered scene, from 20~s wet speech signals, which are obtained similarly to what is described in \ref{sec:datasets}.
% where 20 seconds of speech is convolved with a single SRIR to obtain a {wet speech} that is given to the speaker embedding extractor.
We choose utterances that have not been used in the mixture computation, as well as an SRIR from another room.
% More enrollment embeddings can be added to the enrollment pool. 
We run experiments with varying number of enrollment embeddings $\mathbf{M=2, 10, 20, 30}$.

For the fragment-related embeddings, either the whole duration, or the fragment beginning is used for embedding extraction. {We perform experiments using the fragment's beginning to go towards a low latency system, where track identity could be handled using speaker embeddings at each fragment appearance.}
We consider using the first \textbf{250, 500, 750, 1000, 1500~ms} of the fragments. We denote as \textit{whole} the case where entire fragments lengths are used.

\subsection{Beamforming} 
% \subsubsection{Beamformers}
The fragment-related embedding quality can be impacted by the beamformer used~\cite{dowerahHowLeverageDNNbased2022}.
We experiment using:
\begin{enumerate*}
    \item An ideal beamformer (\textbf{Ideal})\footnote{In this paper, we use the term \textit{ideal beamformer} to refer to the wet signals~$\*x_j$ before the mixture computation.}. 
    \item A FOA Delay-and-Sum (\textbf{DS})~\cite{baqueAnalyseSceneSonore2017} beamformer.
    \item A Minimum Variance Distortionless Response (\textbf{MVDR}) beamformer~\cite{caponHighresolutionFrequencywavenumberSpectrum1969}.
    % \item The Generalized Eigenvalue Decomposition approximation of the Multichannel Wiener Filter (GEVD-MWF)~\cite{serizelLowrankApproximationBased2014} denoted later as \textbf{GEVD}.
\end{enumerate*} 
% The spatial covariance matrices used to compute the MVDR beamformer weights are obtained with time-frequency masks, estimated using a U-net neural network architecture~\cite{boscaDilatedUnetBased2021}.
{To estimate the noise spatial covariance matrix needed to compute the MVDR weights, we use the pretrained neural network of~\citet{boscaDilatedUnetBased2021}, that estimates time-frequency masks, given FOA signal and DoA information.
}

\begin{figure}[!tbp]
    \centering
    \includegraphics[width=0.95\linewidth]{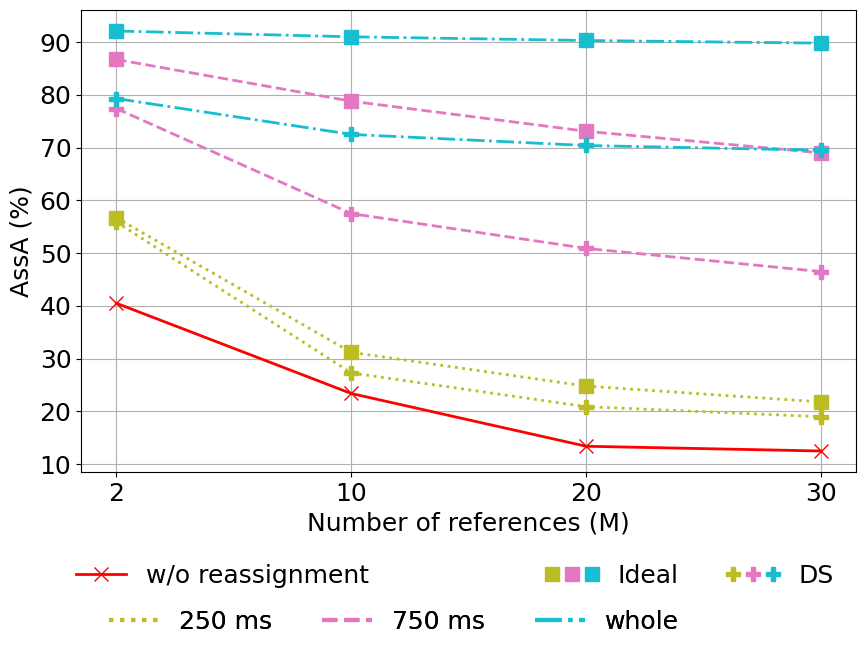}
    \caption{Impact of the number of enrollments $M$ on the reassignment system AssA score for three 
    fragment input durations and two beamformers. 
    % Dataset: $60-180\degree$ - Tracker: GT
    }
    \label{fig:number_enrollments}
\end{figure}

\subsection{Evaluation Metrics}
\textcolor{red}{We propose to use metrics measuring track identity assignment performance, namely the tracking association accuracy metric (\textbf{AssA}) [\%] taken from the multi-object tracking community~\cite{luitenHOTAHigherOrder2021}, and adapted to DoA tracking~\cite{iatarieneTrackingIntermittentMoving2025}, as well as the LOCATA~\cite{eversLOCATAChallengeAcoustic2020} Track Swap Rate (\textbf{TSR}) and Track Fragmentation Rate (\textbf{TFR}) [$s^{-1}$] metrics.
% introduced by~\citet{luitenHOTAHigherOrder2021} 
% from the multi-object tracking community~\cite{luitenHOTAHigherOrder2021}, and adapted to DoA tracking~\cite{iatarieneTrackingIntermittentMoving2025b}. 
AssA evaluates the consistency of entire predicted trajectories against ground truths, in contrast with the LOCATA challenge metrics, which compares ground truths and predictions at the frame level~\cite{iatarieneTrackingIntermittentMoving2025}.}

% To compare tracking systems, we measure identity related metrics as well as spatial metrics. In addition to the global AssA score, we measure the LOCATA~\cite{eversLOCATAChallengeAcoustic2020} frame-level identity metrics Track Swap Rate (\textbf{TSR}) and Track Fragmentation Rate (\textbf{TFR}) [$s^{-1}$]~\cite{iatarieneTrackingIntermittentMoving2025b}.
For the spatial performance evaluation, we measure the Localization Error (\textbf{LE}) [°] as the mean angular distance between ground truth and predicted DoAs~\cite{perotinCRNNBasedMultipleDoA2019}.

\section{Results}
\label{sec:res}

We start by assessing the utility of the proposed fragment-level embedding-based reassignment system to improve tracking of intermittent and moving speakers in \ref{sec:res_1}. We study the factors impacting embedding extraction and thus the entire system (beamforming and fragment input duration) in \ref{sec:res_2}. We finally analyze the impact of spatial trajectory quality by comparing in \ref{sec:res_3} the performances of several trackers.

We run experiments with a $80-20\%$ bootstrap and mean metric scores are displayed. Standard deviation did not exceed~1\% in all the experiments.
Fig. \ref{fig:number_enrollments} displays the impact of increasing the number of enrollments for the Ideal and DS beamformers\footnote{We choose not to display the results using MVDR beamformer in Fig. \ref{fig:number_enrollments}, since they are similar to the ones using the DS beamformer.}, for three fragment input durations (250~ms, 750~ms and \textit{whole}). 
Fig \ref{fig:inputduration} displays the evolution of the AssA score when increasing the fragment input duration for the three beamformers (Ideal, DS, MVDR).
The AssA score before reassignment is displayed in red. For both figures, the tracker GT is used on the $60-180\degree$ dataset.

\subsection{{Identity reassignment system global performances}}
\label{sec:res_1}

% \subsubsection{Usefulness of speaker embeddings}

From both figures, we can see that for any beamformer, input duration or number of enrollments, the proposed reassignment system improves performances over the baseline red scores before reassignment. 
This assesses the usefulness of speaker embeddings as identity-related observations to enhance tracking of intermittent and moving speakers.

% \subsubsection{Impact of number of enrollments $M$}

As illustrated by Fig. \ref{fig:number_enrollments}, increasing the number of enrollments $M$ has a negative impact on the AssA score before reassignment, going from 40.5\% when $M=2$ to 12.5\% when $M=30$. 
When given ideal conditions for fragment-level speaker embedding extraction , i.e, \textit{whole} input durations and Ideal beamforming (blue curve with square marker), our proposed system becomes resilient to increasing $M$. 
% However, it becomes more and more affected by it as the input duration decreases, as shown by the pink curves (750~ms) then the green curves (250~ms).

\begingroup
\centering
\setlength{\tabcolsep}{12pt} % Default value: 6pt
\renewcommand{\arraystretch}{1} % Default value: 1
\begin{table}[!tbp]
\centering
\hspace*{-1cm}
\caption{Impact of speaker proximity on the AssA score for the three beamformers and three fragment input durations. Tracker: GT - Number of enrollments $M=J=2$.}
\label{tab:close_speakers}
\begin{tabu}{ccccc}
\hline
\multirow{2}{*}{Beamformer} &   \multirow{2}{*}{Dataset} & \multicolumn{3}{c}{Fragment duration [ms]} \\ \cline{3-5} 
  &  & 250 & 750 & \textit{whole} \\ \hline
  \rowfont{\color{gray}}
\multirow{2}{*}{Ideal}  & 
     $25-60\degree$  & 55.8 & 86.9 &	93.0 \\ 
\rowfont{\color{gray}}
  & {$60-180\degree$} & 56.7 & 86.7 &	92.1    \\ \hline
   
\multirow{2}{*}{DS}  & 
   $25-60\degree$ & 52.1 &	71.4 &	67.6  \\ 
    & $60-180\degree$ & \textbf{55.9} &	\textbf{77.4} &	\textbf{79.3}  \\ \hline
\multirow{2}{*}{MVDR} & 
   $25-60\degree$ &  52.9 & 72.2 &	72.1  \\ 
   &  $60-180\degree$ &  \textbf{55.5} &	\textbf{78.7} &	\textbf{81.6}   \\ 
\end{tabu}
\end{table}
\endgroup

\subsection{Impact of fragment-level speaker embedding quality}
\label{sec:res_2}

% \subsubsection{Input duration for embedding extraction}

\begin{figure}[!tbp]
    \centering
    \includegraphics[width=0.95\linewidth]{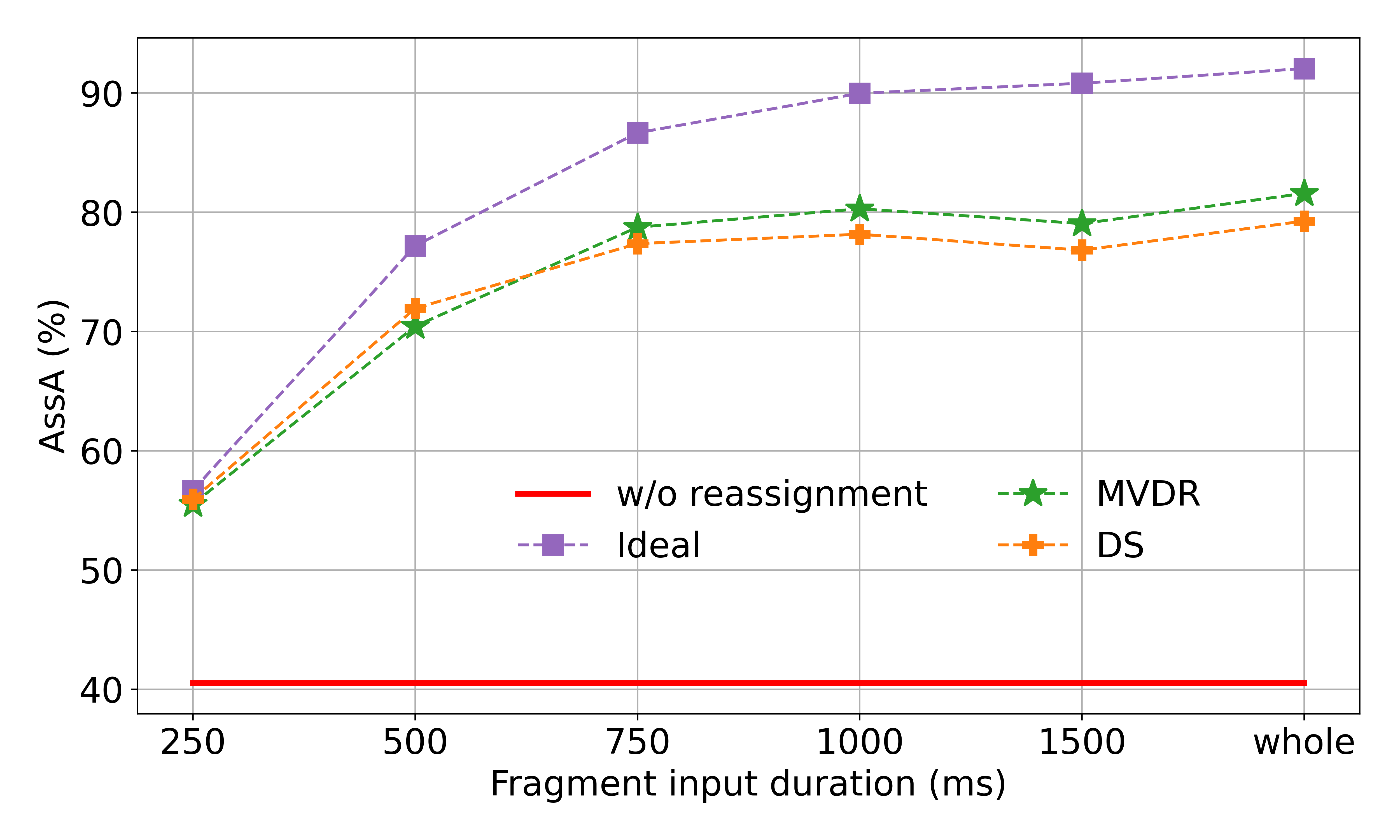}
    \caption{\footnotesize Impact of the fragment input duration on the reassignment system AssA score {for three beamformers} 
    % - Dataset: $60-180\degree$ - Tracker: GT 
    - Number of enrollments $M=J=2$}
    \label{fig:inputduration}
\end{figure}

Speaker embedding quality depends on input signal duration for extraction~\cite{cui_improving_2024}, and on separation quality in the presence of overlapping speakers~\cite{dowerahHowLeverageDNNbased2022}. 
% as it is our case. 
As illustrated by Fig \ref{fig:inputduration}, decreasing the input duration has a negative impact on the reassignment system performances. For the Ideal beamformer, it goes from~92.1\% (\textit{whole}) to~56.7\% (250~ms).
% This illustrates the degradation of speaker embedding extraction when given short input durations.

As for beamforming, Fig. \ref{fig:inputduration} shows that using the MVDR beamformer leads to better reassignment performances than using the DS. This emphasizes the need for robust separation for embedding extraction. 
{Fig. \ref{fig:inputduration} also shows that regardless of the beamformer, performances severely degrades on the shortest input durations. 
This shows the limit of the pre-trained speaker embedding model at extracting meaningful speaker representations from such short context.
}
% RS: "Regardless of the beamformer, performance severly degrades on short fragments indicating the need for more robust processing in this particular case"
% the impact of separation becomes less and less important as input durations becomes shorter. 

Another factor impacting beamforming quality is the spatial placement of speakers. Table \ref{tab:close_speakers} displays AssA scores for the two datasets of distant ($60-180\degree$) and close ($25-60\degree$) speakers. For both beamformers, better performances are achieved with the distant speakers dataset. 
The DS beamformer is also more impacted by the speaker's spatial distance. Given the \textit{whole} input duration, a drop of 11.7\% can be measured for the DS beamformer when having closer speakers, while it is of 9.5\% for the MVDR beamformer.
 
\begingroup
    \centering
    \setlength{\tabcolsep}{8pt} % Default value: 6pt
    \renewcommand{\arraystretch}{1} % Default value: 1
    \begin{table}[!htbp]
        \centering
        \caption{Spatial tracker performances without reassignment}
        \label{tab:spatial_tracker_wo_reassignement}
            \begin{tabu}{ccccc}
            \hline
            {Tracker} & {LE [°]}  & {TSR [$s^{-1}$]} & {TFR [$s^{-1}$]}
            & {AssA [\%]} \\ \hline
             \rowfont{\color{gray}}
            {GT} & {0.43} & {0.16} &  {0.64} & {40.5} \\ \hline
            {EST} & {\textbf{6.45}} & {\textbf{0.30}} & {0.93} & {36.7} \\ \hline
            {NN} & {9.43}  & {0.87} & {\textbf{0.87}} & {\textbf{38.1}} \\
            \end{tabu}
    \end{table}
    \endgroup
    
\begingroup
    \centering
    \setlength{\tabcolsep}{12pt} % Default value: 6pt
    \renewcommand{\arraystretch}{1} % Default value: 1
    \begin{table}[!htbp]
        \centering
        \caption{Spatial tracker performances (AssA score [\%]) with reassignment for two beamformers and three input durations - Number of enrollments $M=J=2$}
        \label{tab:spatial_tracker_with_reassignment}
        \begin{tabu}{ccccc}
            \hline
           \multirow{2}{*} {Tracker} & \multirow{2}{*}{Beamformer} & \multicolumn{3}{c}{Fragment duration [ms]} \\ \cline{3-5} 
             &  & {250} & {750} & {\textit{whole}} \\ \hline
             % &  &  &  &  \\ \hline
             \rowfont{\color{gray}}
            \multirow{2}{*}{GT} & Ideal & 56.7  & 86.7 & 92.1  \\
            \rowfont{\color{gray}}
             & DS & 55.9  & 77.4  & 79.3  \\ \hline
            \multirow{2}{*}{EST} & Ideal & \textbf{53.1} & \textbf{79.0} & \textbf{82.1} \\
             & DS & \textbf{50.3} & \textbf{71.2} & \textbf{74.4} \\ \hline
            \multirow{2}{*}{NN} & Ideal & 51.6  & 72.9  & 76.6  \\
             & DS & 46.3  & 63.3  & 66.5
             \end{tabu}
     \end{table}
    \endgroup
\subsection{Impact of the tracking system}
\label{sec:res_3}

Previous results were obtained using the tracker GT to focus on factors impacting fragment-related speaker embedding extraction, given robust fragments of trajectories. 
However, spatial trajectories can have an impact on fragment quality and thus on the reassignment system.
Table \ref{tab:spatial_tracker_wo_reassignement} and Table \ref{tab:spatial_tracker_with_reassignment} display the performances of the three spatial trackers (GT, EST, NN) without and with reassignment {(for 250 ms, 750 ms and \textit{whole} input durations)}, respectively.

From Table \ref{tab:spatial_tracker_wo_reassignement}, we notice that EST is better than NN.
% we can see that the Localization Error (LE) is smaller for EST. This is due to the DoA discretization grid. EST benefits from an additional smoothing thanks to second tracking step.
We notice in particular that the TSR score is better for EST, and shows a rate of 0.87 swaps per second for NN.
Since NN has been trained using Permutation Invariant Training (PIT), it probably suffers from the block permutation problem defined in speech separation community~\cite{chenContinuousSpeechSeparation2020} and in speaker diarization community~\cite{kinoshitaIntegratingEndEndNeural2021}.
This leads NN to predict potentially permuted outputs from one {inference run} to another.
LE score is lower for EST, while TFR score is similar for both EST and NN.
The AssA score is 1.4\% better for NN than it is for EST.

Table \ref{tab:spatial_tracker_with_reassignment} finally illustrates that better spatial trajectories lead to better reassignment performances, since reassignment using the EST tracker performs better for both Ideal and DS beamformer.
Given \textit{whole} input durations, EST tracker is approx. 7.9\% better with the DS beamformer and 5.5\% better with the Ideal beamformer.

\section{Conclusion}
\label{sec:conclu}
In this paper, we investigated the use of speaker embeddings as identity-related observation for tracking, and addressed the problem of tracking intermittent and moving speakers. 
% We proposed approach relying on speaker embeddings to improve the identity assignment of tracking systems in such scenarios.
We proposed a {baseline} solution that performs fragment-level identity reassignment {in a post tracking step, given an enrollment pool}. This method involves beamforming and speaker embedding extraction through a pretrained general model.
% {To do so, fragment-related speaker embeddings are obtained using a pre-trained model, that is applied over a beamformed signal obtained using the fragment's spatial information.} 
% Fragment-level identity reassignment is then performed by comparing these speaker embeddings to a pool of enrollment embeddings.
Experimental results showed that such 
% fragment-level identity reassignment
method improves significantly the identity assignment performance of several tracking systems, {assessing the usefulness of speaker embeddings to complement DoAs for speaker tracking}. 
% While still being superior to no reassignment at all, 
However, performances were affected by beamforming, input duration for embedding extraction, and by the spatial trajectory quality. 
% The longer the fragment, the higher the number of DoA errors + swaps mistakes taken into account (both are worse for NN tracker so...).
% \textbf{This underlines the need to be able to compute minimal context speaker embeddings}
This illustrates the unsuitability of such general pretrained models to extract robust speaker embeddings, {given short signals containing interfering speakers.}
% This motivates future studies towards the design of more {context-aware} speaker embedding extraction models.
% , able to retrieve discriminative speaker representations from short inputs and despite interfering speakers.
% Having fragmented spatial trajectories also lead to lower performances.

% Specifically, using longer fragments leads to higher-quality speaker representations and, consequently, better assignment performance. We also compared different beamformers, with the MVDR beamformer outperforming others when having spaced speakers and when the input duration for embedding extraction exceeds 500~ms.
% {Finally, using estimated DoAs resulted in lower performance than using ground truth DoAs, but it still outperformed the DoA-based tracker without embedding, especially when using long fragments input duration. 
% Future studies can focus more on the impact of estimated DoAs, by studying the impact of localization errors and short and fragmented tracks, and designing more robust ID assignment methods.}
% with shorter segments for embedding extraction,
% As the tracker system generates discontinuous, small fragments. Extraction from these small fragments can degrade speech quality.
% However, using longer segments resolves this issue, providing more stable speaker representations.   
% {Future studies can focus on handling short fragmented tracks
% }

% \bibliographystyle{IEEEtran}
% \bibliography{enrollments}
\printbibliography

\end{document}